# Observation of acoustic spin

Chengzhi Shi[1†], Rongkuo Zhao[1†], Yang Long[2†], Sui Yang[1], Yuan Wang[1], Hong Chen[2], Jie Ren[2♦], Xiang Zhang[1,3*]

[1]NSF Nano-scale Science and Engineering Center (NSEC), University of California, Berkeley, 3112 Etcheverry Hall, Berkeley, CA 94720, USA

[2]Center for Phononics and Thermal Energy Science, China-EU Joint Center for Nanophononics, Shanghai Key Laboratory of Special Artificial Microstructure Materials and Technology, School of Physics Sciences and Engineering, Tongji University, Shanghai 200092, China

[3]Materials Science Division, Lawrence Berkeley National Laboratory, 1 Cyclotron Road, Berkeley, CA 94720, USA

[†]These authors contributed equally to this work.

[♦]*Email: xonics@tongji.edu.cn*

[*]*Email: xiang@berkeley.edu*

**Unlike optical waves, acoustic waves in fluids are described by scalar pressure fields, and therefore are considered spinless. Here, we demonstrate experimentally the existence of spin in acoustics. In the interference of two acoustic waves propagating perpendicularly to each other, we observed the spin angular momentum in free space as a result of the rotation of local particle velocity. We successfully measured the acoustic spin, and spin induced torque acting on a lossy acoustic meta-atom that results from absorption of the spin**

**angular momentum. The acoustic spin is also observed in the evanescent field of a guided mode traveling along a metamaterial waveguide. We found spin-momentum locking in acoustic waves whose propagation direction is determined by the sign of spin. The observed acoustic spin could open a new door in acoustics and their applications for the control of wave propagation and particle rotation.**

The spin angular momentum describes the rotation of a vector field [1, 2]. It provides an extra degree of freedom for the control of wave propagation and wave matter interactions. The spin angular momentum of light is a result of the rotation of electric polarization [3]. In addition to the longitudinal spin represented by circularly polarized light where the axis of rotation is parallel to the propagation direction [4], the recently studied optical transverse spin [5] with the axis of rotation perpendicular to the direction of propagation has shown interesting physics such as strong spin orbital interaction and quantum spin Hall effect [6-13].

In fluids such as air and water, because the acoustic wave can be deterministically described by the scalar pressure field [14], the spin degree of freedom in acoustics has not been explored [15-17]. Recent studies have raised the question if acoustic spin can ever exist [15-17]. Similar to optical spin, one may consider acoustic spin as the rotation of the wave polarization given by its local particle velocity, but an acoustic plane wave propagating in free space is a longitudinal wave whose particle velocity always oscillates along the propagation direction and does not rotates [14]. Note that the spin angular momentum is different from the orbital angular momentum observed in acoustic vortices representing the circulation of energy flux [15-17] or helical shaped acoustic or optical beams associated with the twisted wavefront [18-24].

Here, we report the existence of spin angular momentum in airborne acoustics characterized by the rotation of local particle velocity. A spinning local particle velocity $\vec{v}$ can be decomposed into two perpendicular components $v_x$ and $v_y$ that are the same in amplitude but with 90 degrees difference in phase. The local particle velocity rotates clockwise or counterclockwise circularly depending on the relative phase difference (Figure 1a). For convenience, we define the clockwise or counterclockwise acoustic spin as spin up or spin down, respectively. This rotating particle velocity field can be observed in the interference of two beams with equal amplitudes propagating perpendicularly to each other (Figure 1b). Each beam contributes a component to the particle velocity field ($v_x$ or $v_y$) in the interference pattern. The phase difference between these two orthogonal components is determined by the position, resulting in spin up or spin down region respectively.

To quantify the strength of the acoustic spin, we define the angular momentum carried by the spinning acoustic field in a unit volume as $\vec{s} = \text{Im}(\rho_0 \vec{v}^* \times \vec{v})/2\omega$, where $\rho_0$ is the density of air and $\omega$ is the frequency of the acoustic field, which is derived from the acoustic angular momentum separating from the orbital angular momentum

(Supplementary Materials). We refer $\vec{s}$ as spin density. The nonzero cross-product of the complex conjugate particle velocity with itself characterizes the rotation of the particle velocity field. For a circularly rotating particle velocity field, the spin density reaches its maximum value and represents the strongest angular momentum. For a linearly oscillating velocity field, both the spin density and the angular momentum are zero.

In experiment, two acoustic beams are excited by two speakers placed at two neighboring sides of the setup (Figure 2a). By measuring the time dependent pressure field $p$ (Figure 2d), the local particle velocity is given by $\vec{v} = -\nabla p/i\omega\rho_0$, where $i$ is the imaginary unit. We found that the velocity rotates clockwise at the center, resulting in a spin up acoustic field (Figure 2d and Supplementary Material Movies 1 and 2). The measured pressure field and spin density shown in Figures 2d and 2e agrees with our simulations (Figures 2b and 2c). On the contrary, in the acoustic field excited only by one speaker, no spin is observed (Figures 2f and 2g).

The acoustic spin carries angular momentum, which can induce a torque through spin matter interaction. In our study, an acoustic meta-atom that can support a dipole resonance – with the air coming out from half of the meta-atom and flowing into another half at a certain moment – is placed in a spinning particle velocity field. Because the meta-atom is lossy due to the mechanical deformation of the meta-atom and the viscosity of air when the gas goes in and out through the tiny slits, the excitation of dipole moment is always slightly delayed in phase comparing to the exciting velocity field [25], which means that the excited dipole moment is not parallel with the exciting velocity field. Because an acoustic dipole tends to align with the velocity field, the misalignment drives the meta-atom, which provides torque acting on it. In other words, the meta-atom obtains angular momentum from the acoustic waves by absorbing the spin angular momentum.

To measure the spin induced torque, we made a coiled space meta-atom (the inset of Figure 2h and Supplementary Materials Section II), which supports a dipole resonance at 870 Hz. The meta-atom is designed to be symmetric with a cylindrical shape to eliminate any possible torque due to the geometry itself. It is designed to be subwavelength in diameter so as to represent a probe to interact with the local spin. The meta-atom is hanged by a thin copper wire at the center of the interference fields (Figure 2a). A mirror is attached on the meta-atom to reflect a laser beam onto a ruler, which converts the rotation of the meta-atom into the deviation of laser spot. The value of the torque is obtained by multiplying the torsional spring constant with the measured rotation angle (Supplementary Materials Section III).

The torques induced by spin up and spin down acoustic waves are of equal amplitude along $-z$ and $+z$ directions, respectively (Figure 2h). The measured torques follow a quadratic relation with the amplitude of the input voltage loaded on the speakers, which shows that the spin induced torque is linearly proportional to the spin density as predicted (Supplementary Materials Section V). The noise is mainly contributed by environmental random vibrations. As a control experiment, the torque from one acoustic beam alone is undetectable, which means that a single beam without interference does not carry angular momentum. Note that the torque in this work originated from the delayed dipole resonance is fundamentally different from the acoustic viscous torque resulted from rotating acoustic particle velocity field in viscous fluids [26-29]. The estimated viscous torque based on the theory described in [27] is one order smaller than the measured torque shown in Figure 2b. The torque observed here can exist in non-viscous fluids as long as the meta-atom itself is lossy and can be polarized by the particle velocity field.

Spin-momentum locking is one of the most interesting physics used for chiral quantum circulators and asymmetric wave transports [6-13] in optics. We also experimentally observed acoustic spin-momentum locking phenomenon. In the acoustic waves supported by a metamaterial waveguide composed of periodic grooves, the evanescent field propagates along the waveguide but decays exponentially in the perpendicular direction (Figure 3a). The two components of the particle velocity satisfy $v_x = \pm ik v_y/\tau$, where $k$ is the wave number along the waveguide and $\tau$ is the decaying constant in the perpendicular direction (Supplementary Materials Section VI). Therefore, the *x* and *y* components of the particle velocity are innately 90 degrees out-of-phase everywhere. The sign determines that the propagation direction is locked with the spin direction. The spin up wave propagates to the right and the spin down wave to the left.

The acoustic spin-momentum locking is demonstrated experimentally with the metamaterial waveguide confined by two rigid walls (Figure 3b). Four mini-speakers are packed together and mounted in the vicinity of the metamaterial waveguide with their phases modulated 90 degrees relative to each other to mimic a rotating acoustic dipole source (Supplementary Materials Section VII). When this dipole source rotates counterclockwise, it excites the spin down acoustic wave which propagates only towards $-x$ direction (Figures 3c (simulation) and 3e (experiment)). On the other hand, the clockwise rotating dipole source excites the spin up acoustic wave propagating only towards $+x$ direction (Figure 3g). The particle velocity fields in Supplementary Material Movies 3 and 4 also show that the acoustic wave with counterclockwise/clockwise rotating particle velocity propagates towards $-x/+x$ direction. The calculated and measured spin densities further confirm the spin-momentum locking phenomenon (Figures 3d, 3f and 3h).

In conclusion, we have demonstrated the existence of spin in acoustics. The spin induced torque is measured to be proportional to the spin density and the spin-momentum locking is shown. The observation of acoustic transverse spin provides the fundamental platform for future studies on acoustic spin physics such as spin orbital interaction, acoustic spin Hall effect, and spin induced non-reciprocal acoustic physics which are important for applications in the control of wave propagation and particle rotation.

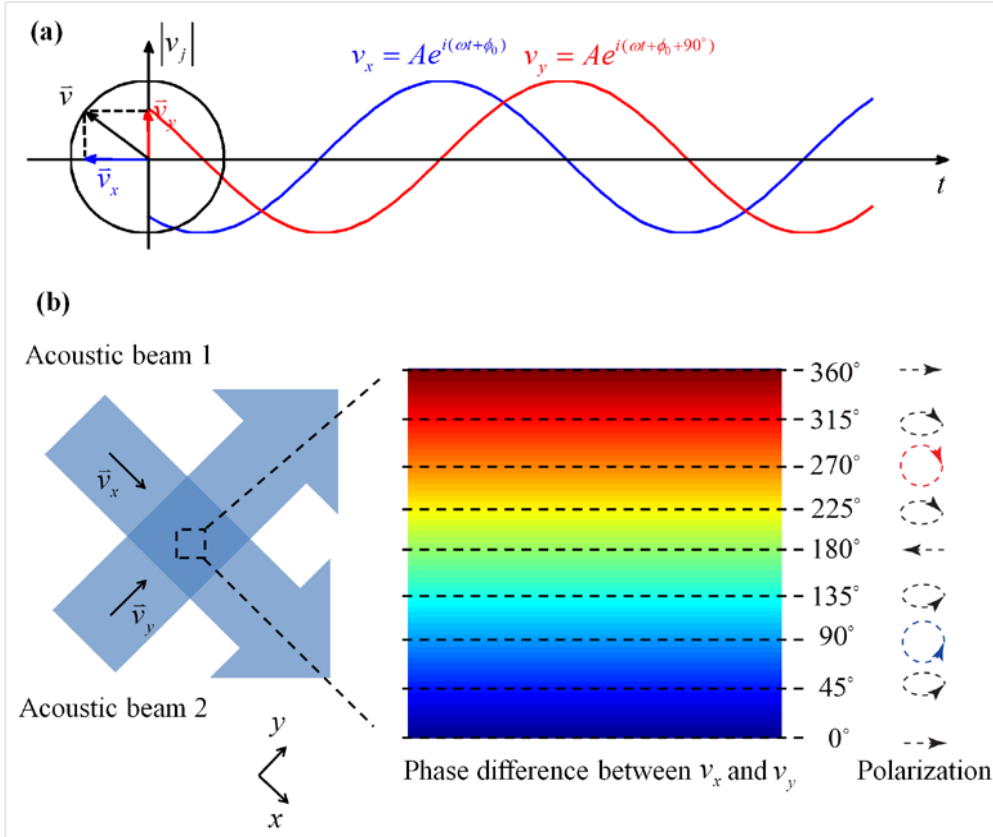

**Figure 1: Acoustic spin as a rotating particle velocity field.** (a) A rotating particle velocity (black arrow) can be decomposed into two components $v_x$ (blue arrow) and $v_y$ (red arrow) along $x$ and $y$ directions. The two components shown as the blue and red lines are 90 degrees out-of-phase. (b) Acoustic spin in the interference of two acoustic beams. Two beams with equal amplitudes propagating along $x$ and $y$ directions contribute $v_x$ and $v_y$ components of the particle velocity field, respectively. The phase difference between $v_x$ and $v_y$ is the function of the position. The zoom-in region shows an area where the phase difference between $v_x$ and $v_y$ changes from 0º to 360º. When the phase difference is 90 or 270 (equivalent to -90) degrees, the local particle velocity field is rotating circularly, resulting in spin up (blue) or spin down (red) acoustic field; When the phase difference is 0, 180, or 360 degrees, the particle velocity field is oscillating along a line. In other cases, the local particle

velocity rotates elliptically.

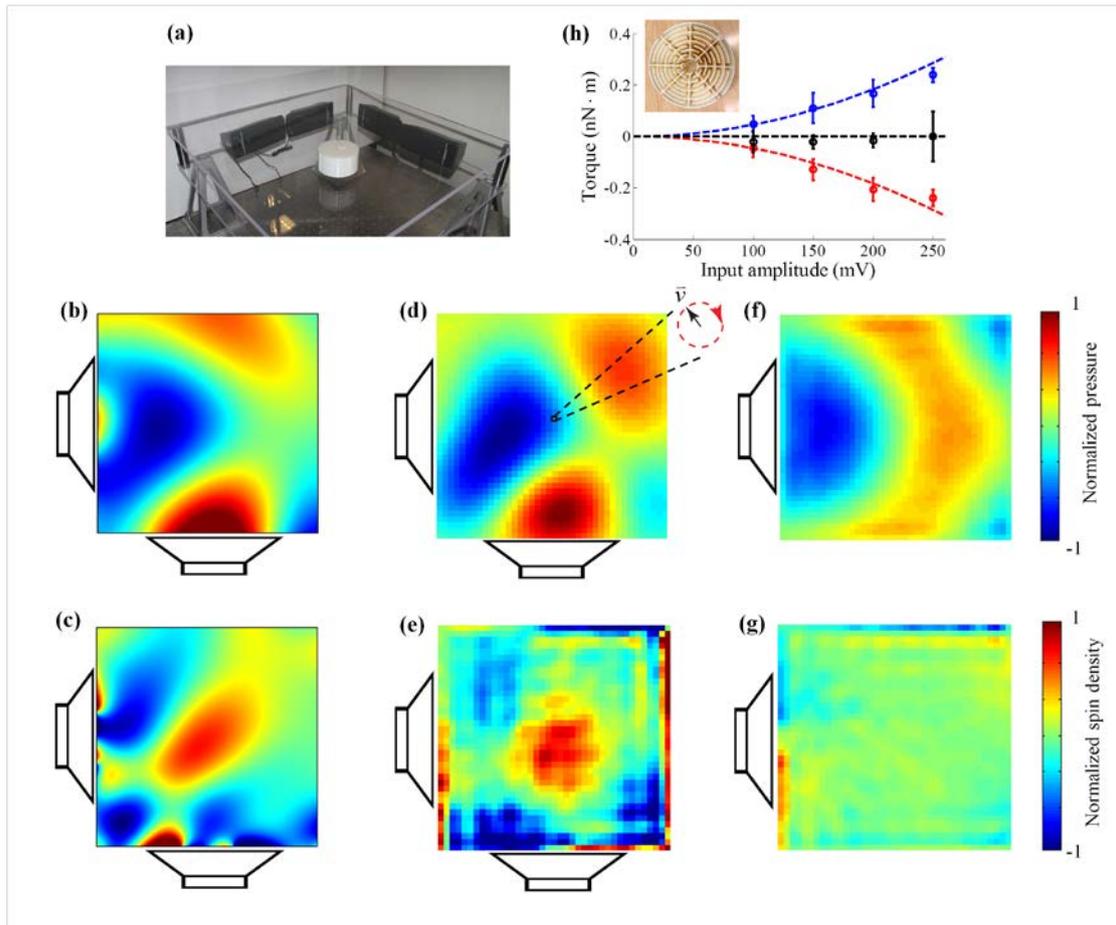

**Figure 2: Experimental observation of acoustic spin.** (a) Experimental setup for the measurement of acoustic spin resulted from the interference of two perpendicular beams and the spin induced torque acting on a coiled space acoustic meta-atom. Two pairs of high power speakers at two neighboring edges emit at 870 Hz with a 90-degree phase difference. The transparent glass walls at the top and the bottom confine acoustic waves propagating at the fundamental mode, mimicking an ideal two-dimensional scenario extended infinitely in the perpendicular direction. The coiled space meta-atom is hanged at the center using a thin copper wire. (b) Simulated and (d) measured pressure fields show a 90-degree phase difference at the center of the interference pattern where the local particle velocity rotates clockwise, resulting in a spin up field. (c) Simulated and (e) measured spin density distributions show the spin density reaches its local maximum at the center where the meta-atom is located. (f) Measured pressure field when only one speaker is on. (g) Measured spin density when only one speaker is on. No spin exists in this case. The measured area for (d, e, f, g) is 40 cm × 40 cm. (h) Measured torques acting on the coiled space meta-atom (inset) versus input voltage amplitude loaded on the speakers. The spin up or spin down acoustic wave applies a negative or positive torque (with respect to $z$ axis in Figure 1b) on the particle. The torques induced by the spin up and spin down waves are equal in amplitude and follow a quadratic relation with the input amplitude, in agreement with our theoretical prediction that the torque is proportional to the spin density.

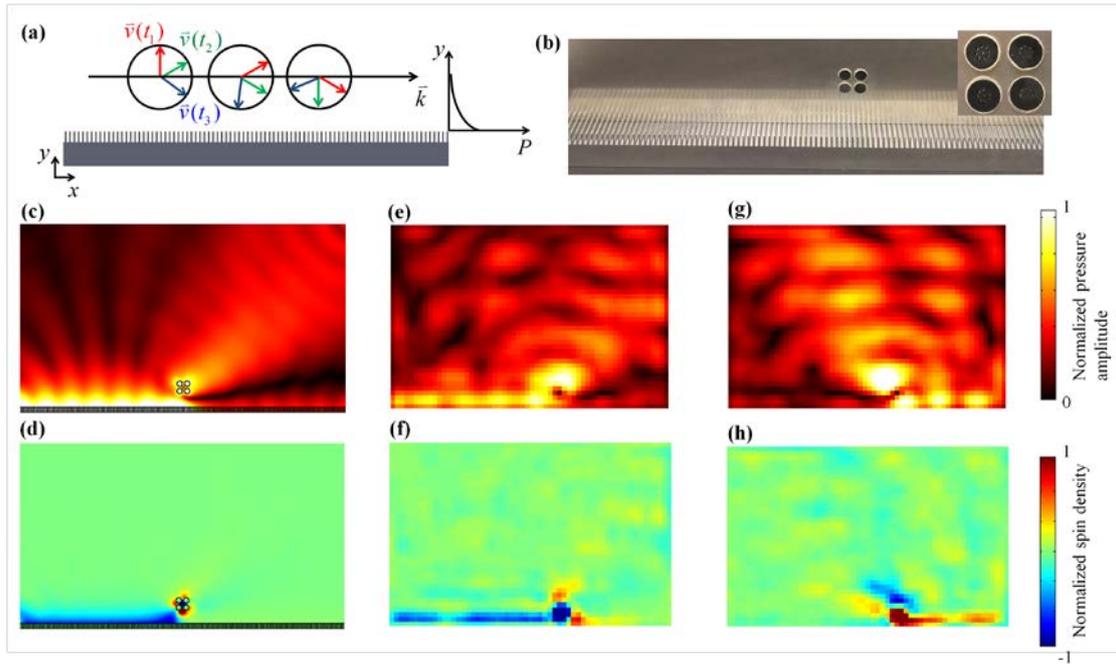

**Figure 3: Spin-momentum locking in acoustics.** (a) Schematic of local particle velocity field of an evanescent acoustic wave supported by a metamaterial waveguide composed of periodic grooves. The acoustic field outside the waveguide decays exponentially along $y$ axis and propagates along $x$ axis. $\vec{v}(t_1)$ (red), $\vec{v}(t_2)$ (green), and $\vec{v}(t_3)$ (blue) represent the particle velocities at time $t_1$, $t_2$, and $t_3$ with $0 < t_1 < t_2 < t_3$, respectively. The local particle velocity rotates clockwise in time. The propagation direction is solely determined by the spin direction, resulting in spin-momentum locking. (b) Experimental setup for demonstrating the spin-momentum locking. Four mini-speakers (inset) are mounted near the acoustic waveguide. A rigid wall on each side confines the acoustic wave propagating in-plane. The four speakers emit at 2 kHz with a 90-degree phase difference between the neighboring speakers, mimicking a rotating acoustic dipole, which excites spin up or spin down acoustic wave determined by the relative phase difference among four speakers. (c, e, g) Normalized amplitudes of simulated and measured pressure fields.

The acoustic metamaterial waveguide is at the bottom of each figure (not shown). The spin down and spin up acoustic waves are excited in (c, e) and (g), respectively. The measured area of (e) and (g) is 60 cm × 40 cm. The spin down acoustic wave propagates only towards left, while the spin up wave propagates towards right, demonstrating the phenomenon of spin-momentum locking. The thermal color scale is used for the pressure amplitude. (d, f, h) Normalized spin density for spin down and spin up acoustic waves in (c), (e) and (g), respectively, confirming the acoustic spin-momentum locking. The jet color scale is used for the spin density.